\documentclass[journal=jacsat,manuscript=article]{achemso}

\usepackage{chemformula}
\usepackage[T1]{fontenc}

\author{Zhi-Bo Ni}
\affiliation[Zhejiang University]
{State Key Laboratory of Silicon and Advanced Semiconductor Materials \&
College of Information Science and Electronic Engineering,
Zhejiang University, Hangzhou 310027, China}
\alsoaffiliation[ZJU-Hangzhou Global Scientific and Technological Innovation Center]
{ZJU-Hangzhou Global Scientific and Technological Innovation Center,
Zhejiang University, Hangzhou 311200, China}

\author{Jiong-Zhao Li}
\affiliation[Department of Chemistry]
{Department of Chemistry,
Zhejiang University, Hangzhou 310027, China}

\author{Jia-Wang Yu}
\affiliation[Zhejiang University]
{State Key Laboratory of Silicon and Advanced Semiconductor Materials \&
College of Information Science and Electronic Engineering,
Zhejiang University, Hangzhou 310027, China}
\alsoaffiliation[ZJU-Hangzhou Global Scientific and Technological Innovation Center]
{ZJU-Hangzhou Global Scientific and Technological Innovation Center,
Zhejiang University, Hangzhou 311200, China}

\author{Xiao-Tian Cheng}
\affiliation[Zhejiang University]
{State Key Laboratory of Silicon and Advanced Semiconductor Materials \&
College of Information Science and Electronic Engineering,
Zhejiang University, Hangzhou 310027, China}
\alsoaffiliation[ZJU-Hangzhou Global Scientific and Technological Innovation Center]
{ZJU-Hangzhou Global Scientific and Technological Innovation Center,
Zhejiang University, Hangzhou 311200, China}

\author{Yun-Ran Wang}
\affiliation[ZJU-Hangzhou Global Scientific and Technological Innovation Center]
{ZJU-Hangzhou Global Scientific and Technological Innovation Center,
Zhejiang University, Hangzhou 311200, China}
\alsoaffiliation[College of Integrated Circuits]
{College of Integrated Circuits,
Zhejiang University, Hangzhou 311200, China}

\author{Dai-Bao Hou}
\affiliation[Zhejiang University]
{State Key Laboratory of Silicon and Advanced Semiconductor Materials \&
College of Information Science and Electronic Engineering,
Zhejiang University, Hangzhou 310027, China}
\alsoaffiliation[ZJU-Hangzhou Global Scientific and Technological Innovation Center]
{ZJU-Hangzhou Global Scientific and Technological Innovation Center,
Zhejiang University, Hangzhou 311200, China}

\author{Yan-Hua Liu}
\affiliation[Zhejiang University]
{State Key Laboratory of Silicon and Advanced Semiconductor Materials \&
College of Information Science and Electronic Engineering,
Zhejiang University, Hangzhou 310027, China}
\alsoaffiliation[ZJU-Hangzhou Global Scientific and Technological Innovation Center]
{ZJU-Hangzhou Global Scientific and Technological Innovation Center,
Zhejiang University, Hangzhou 311200, China}

\author{Wei Fang}
\affiliation[College of Optical Science and Engineering]
{College of Optical Science and Engineering,
Zhejiang University, Hangzhou 310027, China}

\author{Xing Lin}
\affiliation[Zhejiang University]
{State Key Laboratory of Silicon and Advanced Semiconductor Materials \&
College of Information Science and Electronic Engineering,
Zhejiang University, Hangzhou 310027, China}
\alsoaffiliation[ZJU-Hangzhou Global Scientific and Technological Innovation Center]
{ZJU-Hangzhou Global Scientific and Technological Innovation Center,
Zhejiang University, Hangzhou 311200, China}

\author{Chao-Yuan Jin}
\email{jincy@zju.edu.cn}
\affiliation[Zhejiang University]
{State Key Laboratory of Silicon and Advanced Semiconductor Materials \&
College of Information Science and Electronic Engineering,
Zhejiang University, Hangzhou 310027, China}
\alsoaffiliation[ZJU-Hangzhou Global Scientific and Technological Innovation Center]
{ZJU-Hangzhou Global Scientific and Technological Innovation Center,
Zhejiang University, Hangzhou 311200, China}
\alsoaffiliation[College of Integrated Circuits]
{College of Integrated Circuits,
Zhejiang University, Hangzhou 311200, China}

\title[An \textsf{achemso} demo]
  {In-Situ Quantum Optical Measurement for Colloidal Quantum Dots Confined in an Optical Trap}

\abbreviations{IR,NMR,UV}

\begin{document}

\newpage
\begin{abstract}
  While optical manipulation of atomic arrays has reached a high degree of precision and scalability, the stable optical confinement of solution-based artificial atoms like colloidal quantum dots (CQDs) remains hindered by weak trapping forces and thermal fluctuations. High-intensity trapping often compromises the quantum properties of these emitters, creating a significant trade-off between mechanical stability and optical integrity. To overcome this, we propose encapsulating CQDs within a transparent polymer matrix, thereby increasing the effective interaction volume and optical restoring force without altering the emitters themselves. This strategy allows for stable spatial confinement under standard experimental conditions, as evidenced by the direct resolution of positional fluctuations through photoluminescence imaging and trajectory tracking. With averaged position fluctuations below 20 nm, the intrinsic emission spectra and photoluminescence decay dynamics remain largely unaffected, and photon-correlation measurements confirm the full preservation of single-photon emission. These findings establish a robust method for the controlled confinement of colloidal quantum emitters and in-situ quantum-optical measurements for future advancements in quantum-optical manipulation of artificial atoms.
\end{abstract}

\section{1. Introduction}
The rapid progress of quantum information science and technology is fundamentally rooted in the precise preparation, control, and measurement of microscopic quantum systems \cite{gross2021quantum, kaufman2021quantum}. Enabled by advanced optical trapping capabilities, atomic and molecular quantum systems have emerged as one of the most important platforms for quantum information technologies \cite{browaeys2020many, schlosser2001sub, mazzanti2024alignment, lambrecht2017long, vilas2024optical}. Within this landscape, several independent studies have recently demonstrated the simultaneous manipulation and rearrangement of large-scale neutral-atom arrays exceeding 1,000 atoms \cite{manetsch2025tweezer, lin2025ai, gyger2024continuous}, highlighting the unprecedented degree of spatial control and quantum manipulation now achievable in microscopic atomic assemblies.

Extending such optical control to artificial atoms is a compelling objective, as they facilitate the bottom-up assembly of nanoscale solid-state arrays\cite{feynman2011there}. These systems offer distinct advantages in on-chip quantum integration\cite{wan2020large} and hold significant promise for applications in quantum bio-sensing\cite{kairdolf2013semiconductor}. However, compared with atomic and molecular systems, the optical manipulation of artificial atoms remains considerably less established. Preliminary examples have been demonstrated using nanodiamonds containing color centers, where individual particles are confined via optical trapping to probe their embedded quantum properties in solvents and biological environments \cite{horowitz2012electron,geiselmann2013three,russell2018manipulating,russell2021optimizing,stewart2024optical,iyer2024optically}. In these systems, the diamond host particle, typically on the scale of tens to hundreds of nanometers, provides a relatively large physical size for stable optical manipulation in solutions while preserving the quantum properties of the embedded color centers.

Colloidal quantum dots (CQDs) represent another important class of artificial quantum emitters with solution processability, size-tunable optical properties, and broad material tunability. However, their nanoscale dimensions ($\sim$10~nm in diameter) lead to weak optical trapping forces and pronounced Brownian motion, posing substantial challenges for stable spatial confinement and in-situ quantum measurements \cite{bendix2013optical}. Although optical tweezers have been used to trap individual CQDs, effective localization requires relatively high trapping intensities, on the order of $500$ $ \rm mW/\mu m^2$\cite{jauffred2008three, jauffred2010two}. Such strong optical fields introduce a practical trade-off between mechanical confinement and preservation of emitter properties, as increased trapping power may induce heating, photocharging, or other field-induced perturbations\cite{jauffred2014sub, bendix2013optical}. Therefore, stable CQD confinement requires not only sufficient trapping strength but also compatibility between the trapping environment and the optical characteristics of the emitters.

To overcome this limitation, we propose to enhance the trapping strength by expanding the effective optical interaction volume through encapsulation of CQDs in a transparent polymer matrix \cite{negele2013stable, ni2026quantum}. The resulting larger composite particle significantly amplifies the optical restoring force against thermal fluctuations \cite{neuman2004optical}, while fully preserving the intrinsic emission properties of the CQDs. Consequently, stable optical confinement can be readily realized under standard experimental conditions.

By employing this strategy, stable spatial confinement of polymer-encapsulated CQDs has been demonstrated by directly resolving their positional fluctuations using photoluminescence (PL) imaging and trajectory tracking. The intrinsic emission spectra and photoluminescence decay dynamics remain largely unaffected under optical trapping conditions. Furthermore, photon-correlation measurements confirm that single-photon emission is fully preserved under laser illumination. These findings collectively establish a robust strategy for the controlled spatial confinement of colloidal quantum emitters, thereby facilitating advanced optical measurements and future quantum-optical investigations.

\section{2. Results and Discussion}

\subsection{2.1 Enhanced Optical Trapping of Encapsulated CQDs}

\begin{figure}[!ht]
\centering
\includegraphics[width=16cm]{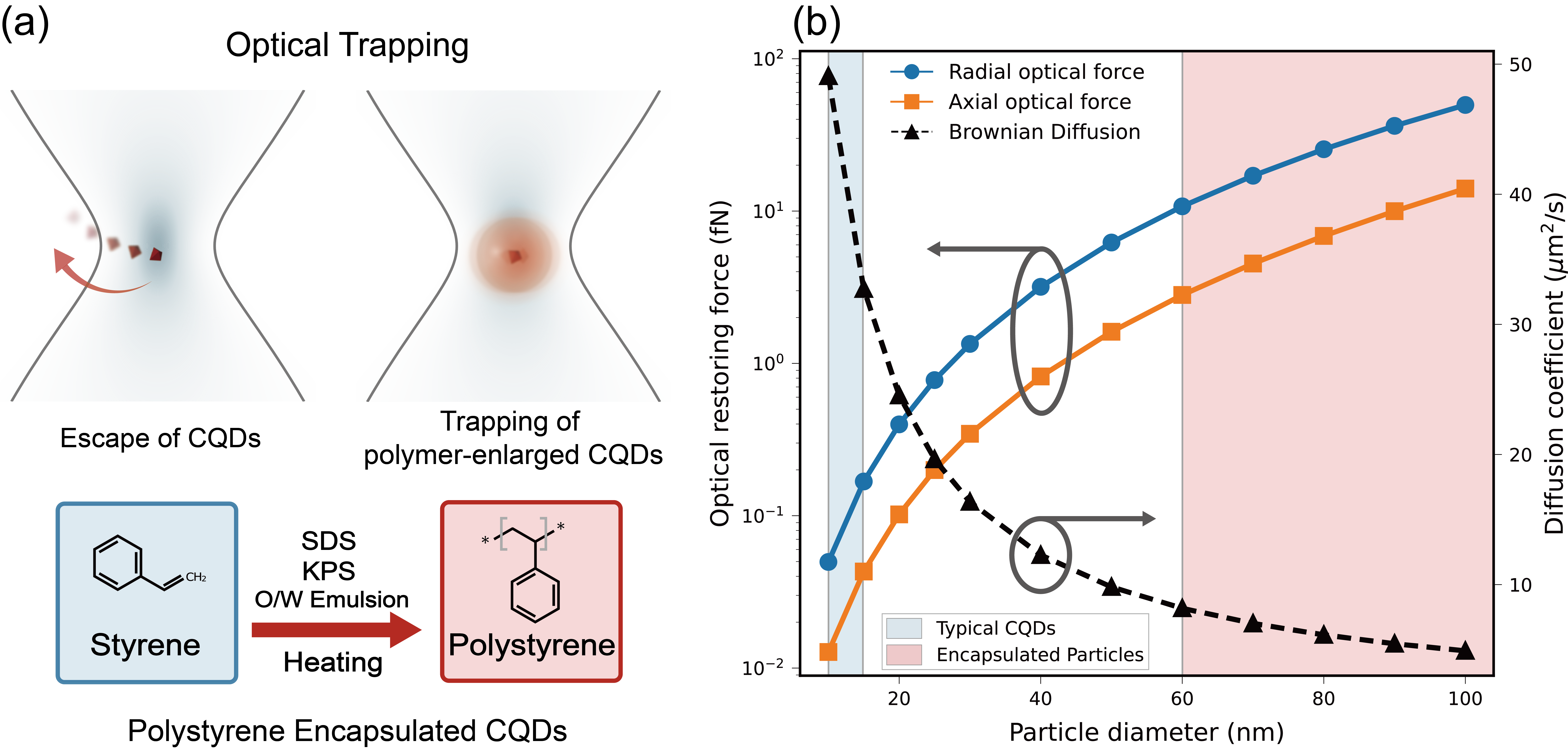}
\caption{Enhanced optical trapping of CQDs through particle-size enlargement by polystyrene encapsulation.
(a) Schematic illustration of the weak optical confinement of small bare CQDs and the enhanced optical trapping after polystyrene encapsulation, together with the encapsulation process via thermal polymerization of styrene.
(b) Calculated radial and axial optical restoring forces and Brownian diffusion coefficient as a function of particle diameter. The shaded regions indicate the representative size ranges of bare CQDs and polymer-encapsulated particles.}
\label{Pic_A}
\end{figure}

Typically, optical tweezers achieve three-dimensional confinement of dielectric particles by tightly focusing a laser beam through a high-numerical-aperture (high-NA) objective, as illustrated in Figure~\ref{Pic_A}(a). The optical force acting on a dielectric particle can be broadly divided into two contributions: the gradient force and the scattering force. In the radial direction, the gradient force drives the particle toward the high-intensity center of the focused beam and provides the dominant confinement. Along the axial direction, the particle experiences both an axial gradient force toward the focal region and a forward scattering force arising from momentum transfer from the incident light. The balance between these two contributions determines the axial trapping position\cite{bustamante2021optical}. In a liquid-phase optical trapping environment, the confined particle is additionally subjected to Brownian motion, which produces stochastic positional fluctuations around the trapping position. Stable trapping therefore requires the optical confinement to remain sufficiently strong against these thermally driven perturbations.

Within the Rayleigh regime, where the particle size is much smaller than the trapping wavelength, both the optical gradient force and optical scattering force exhibit a strong dependence on particle size and the distribution of the electromagnetic field of the optical tweezers\cite{harada1996radiation},

\begin{equation}
F_{\rm grad}
=
\frac{2\pi n_m r^3}{c}
\left(
\frac{m^2-1}{m^2+2}
\right)
\nabla I,
\end{equation}
and

\begin{equation}
F_{\rm scat}
=
\frac{n_m I}{c}\sigma_{\rm scat},
\end{equation}
with the Rayleigh scattering cross section
\begin{equation}
\sigma_{\rm scat}
=
\frac{8\pi}{3}
k_m^4 r^6
\left(
\frac{m^2-1}{m^2+2}
\right)^2.
\end{equation}
Here, $n_m$ is the refractive index of the surrounding medium, $c$ is the speed of light in vacuum, $r$ is the particle radius, $m$ is the relative refractive index between the particle and the surrounding medium, $I$ is the optical intensity, $\nabla I$ is its spatial gradient, and $k_m$ is the wave number in the surrounding medium.

The thermal motion of the particle is likewise size dependent. For a nanoparticle suspended in a viscous medium, the Brownian diffusion coefficient is described by the Stokes--Einstein relation,

\begin{equation}
D = \frac{k_B T}{6\pi\eta r},
\end{equation}
where $\eta$ is the dynamic viscosity of the surrounding medium. Thus, the optical forces and Brownian diffusion follow opposite size-dependent trends, which directly influence the trapping stability. 

To illustrate the influence of particle size on optical trapping, we evaluated the optical forces and Brownian diffusion coefficient as a function of particle diameter. The trapping field was assumed to be an ideal Gaussian beam with a wavelength of 1064~nm and a beam waist of $w_0=0.532~\mu\mathrm{m}$. A peak power density of $50~\mathrm{mW}/\mu\mathrm{m}^2$ at the beam waist was used for the calculations. The radial and axial optical forces were calculated using the Rayleigh-force expressions above and compared with the corresponding Brownian diffusion coefficient. As shown in Figure~\ref{Pic_A}(b), increasing the particle diameter leads to a substantial increase in both the radial and axial optical restoring forces, while the Brownian diffusion coefficient decreases. 

These findings motivate the strategy of increasing the effective particle size to enhance optical confinement. However, directly enlarging the semiconductor nanocrystal itself is often impractical. Expanding the CQD size via continuous epitaxial growth of thick inorganic shells is limited by the structural compatibility between the constituent materials \cite{gong2016strain, shang2017colloidal}. Excessive shell growth can lead to accumulated lattice strain and structural defects, introducing non-radiative recombination pathways that degrade the photoluminescence properties of CQDs \cite{gong2016strain}. Consequently, it is desirable to increase the effective trapping volume without requiring a physical enlargement of the semiconductor nanocrystal itself.

To achieve this, CQDs were encapsulated within a transparent polystyrene matrix via thermal polymerization in an oil-in-water emulsion. In this synthesis, sodium dodecyl sulfate (SDS) serves as the surfactant to stabilize the emulsion, while potassium persulfate (KPS) acts as the radical initiator for styrene polymerization. Following a previously reported protocol \cite{ni2026quantum}, the final particle diameter was controlled by adjusting the relative amounts of SDS and styrene. The resulting polystyrene-encapsulated quantum dots (PQDs) exhibit diameters of up to approximately 100~nm, approximately an order of magnitude larger than those of the pristine CQDs. Crucially, the PS matrix (refractive index $n\approx1.59$) provides a sufficient refractive-index contrast against the surrounding aqueous medium ($n\approx1.33$) to enable robust optical confinement. Furthermore, as corroborated by previous studies \cite{negele2013stable,ni2026quantum}, this polymer encapsulation strategy successfully preserves the single-photon emission characteristics of the embedded CQDs.

\subsection{2.2 Experimental Platform}

\begin{figure}[!ht]
\centering\includegraphics[width=16cm]{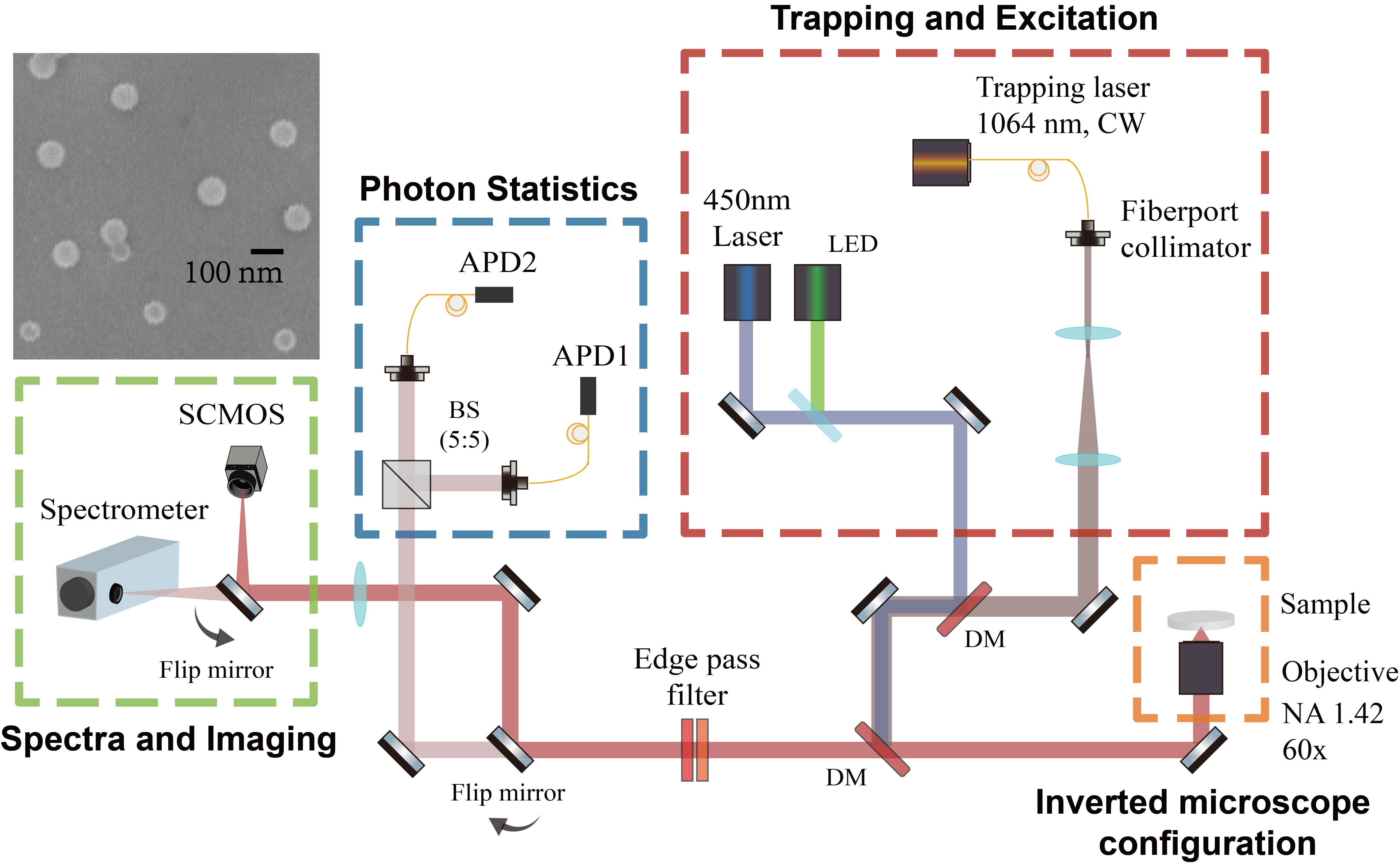}
\caption{Custom-built inverted micro-photoluminescence platform with optical trapping. The system integrates a 1064~nm CW trapping laser, 450~nm excitation sources, a high-NA oil-immersion objective, and switchable detection pathways for spectroscopy, wide-field imaging, and photon correlation measurements.}
\label{Pic_B}
\end{figure}

Optical trapping and characterization were conducted using a custom-built inverted micro-photoluminescence (micro-PL) platform, as depicted in Figure~\ref{Pic_B}. The inset in the upper-left corner shows an SEM image of the PQDs, with corresponding TEM characterization of similar samples reported in Ref.~\cite{ni2026quantum}. This system integrates optical trapping, fluorescence excitation, imaging, spectroscopy, and time-resolved photon detection within a unified optical path. A 1064~nm continuous-wave (CW) fiber laser serves as the trapping source, while the photoluminescence of the CQDs (centered at $\sim$670~nm) is excited using either CW or picosecond pulsed diode lasers operating near 450~nm. A beam-expanding telescope is used to enlarge the incident beam for sufficient filling of the objective back aperture and tighter focusing. Optical trapping, excitation, and fluorescence collection are performed through a single high-numerical-aperture oil-immersion objective (60$\times$, NA~1.42). The collected fluorescence is separated from the trapping and excitation beams using a dichroic mirror and optical filters, and then directed to switchable detection pathways. Spectral measurements are acquired using a blazed grating spectrometer, while an sCMOS camera is employed for imaging and trajectory tracking. Furthermore, time-resolved and photon-correlation measurements are conducted using single-photon avalanche photodiodes (SPADs) integrated with a time-correlated single-photon counting (TCSPC) system.

The high-NA focusing configuration imposes additional requirements on focal-field control because the axial focal position and spatial field distribution are sensitive to wavelength and refractive-index mismatch. First, the wavelength difference between the 670~nm fluorescence and the 1064~nm trapping beam leads to an axial offset between the fluorescence imaging plane and the trapping focus. A slight adjustment of the telescope lens separation changes the beam convergence and thereby compensates for this focal mismatch. Second, even after this wavelength-dependent focal offset is compensated, the 1064~nm trapping field itself is modified by the refractive-index mismatch at the sample interface. Focusing across the glass--water interface introduces spherical aberration and modifies the focal-field distribution. Its influence is evaluated through focal-field reconstruction in the following subsection.

\subsection{2.3 Focal-Field Optimization}

The inverted architecture of the experimental setup and the use of oil-immersion microscopy introduce further complexity to the optimization of optical forces. To ensure the optical-force simulations accurately reflect the experimental conditions, we first reconstructed the three-dimensional vectorial focal field of the 1064~nm trapping beam. In our system, polystyrene encapsulated colloidal quantum dots (PQDs) dispersed in deionized water were confined within a sealed sample cell featuring an etched recess in the glass substrate. The trapping beam was then tightly focused through the glass--water interface using a high-numerical-aperture (NA) oil-immersion objective. The refractive-index mismatch at this interface induces spherical aberration, which can substantially modify the local field intensity and distribution. Consequently, the focal-field reconstruction was performed under the specific experimental geometry rather than assuming ideal focusing conditions.

Using a vectorial Debye model, we reconstructed the focal field by accounting for the glass--water interface \cite{torok1995electromagnetic, torok1997electromagnetic}. The detailed formulation of the focal-field reconstruction is provided in the Supporting Information. Propagation through the refractive-index-mismatched interface shifts the actual focus from the preset geometrical position and alters the focal distribution, with the nominal focal depth defined as the preset focus measured from the glass--water interface into the aqueous region. The reconstructed focal field serves two primary purposes. First, it allows us to determine suitable focusing conditions that preserve sufficient trapping intensity. Second, on the basis of this field, we calculate the trapping potential to evaluate whether the resulting optical confinement is strong enough to overcome Brownian fluctuations and maintain stable nanoparticle trapping.

The focal-depth dependence of the reconstructed trapping field is shown in Figure~S2 of the Supporting Information. The focal-field widths are defined using the $1/e^2$ intensity criterion, with $2w_0$ representing the radial width and the corresponding axial width defined along the beam-propagation direction. As the nominal focal depth increases from $0.5~\mu\mathrm{m}$ to $10~\mu\mathrm{m}$, the axial width increases by approximately a factor of two, while the peak intensity decreases by approximately 50\%, normalized to that at a focal depth of $0.5~\mu\mathrm{m}$. These results indicate that increasing the focal depth progressively broadens the focal field and reduces its peak intensity. Since the optical gradient force depends on the spatial variation of the intensity, both effects lead to weaker optical confinement, particularly along the axial direction. A nominal trapping depth below approximately $2~\mu\mathrm{m}$ is preferable for maintaining a sufficiently confined trapping field. Accordingly, a trapping depth of $2~\mu\mathrm{m}$ was selected for the subsequent calculations.

\begin{figure}[!ht]
\centering
\includegraphics[width=16cm]{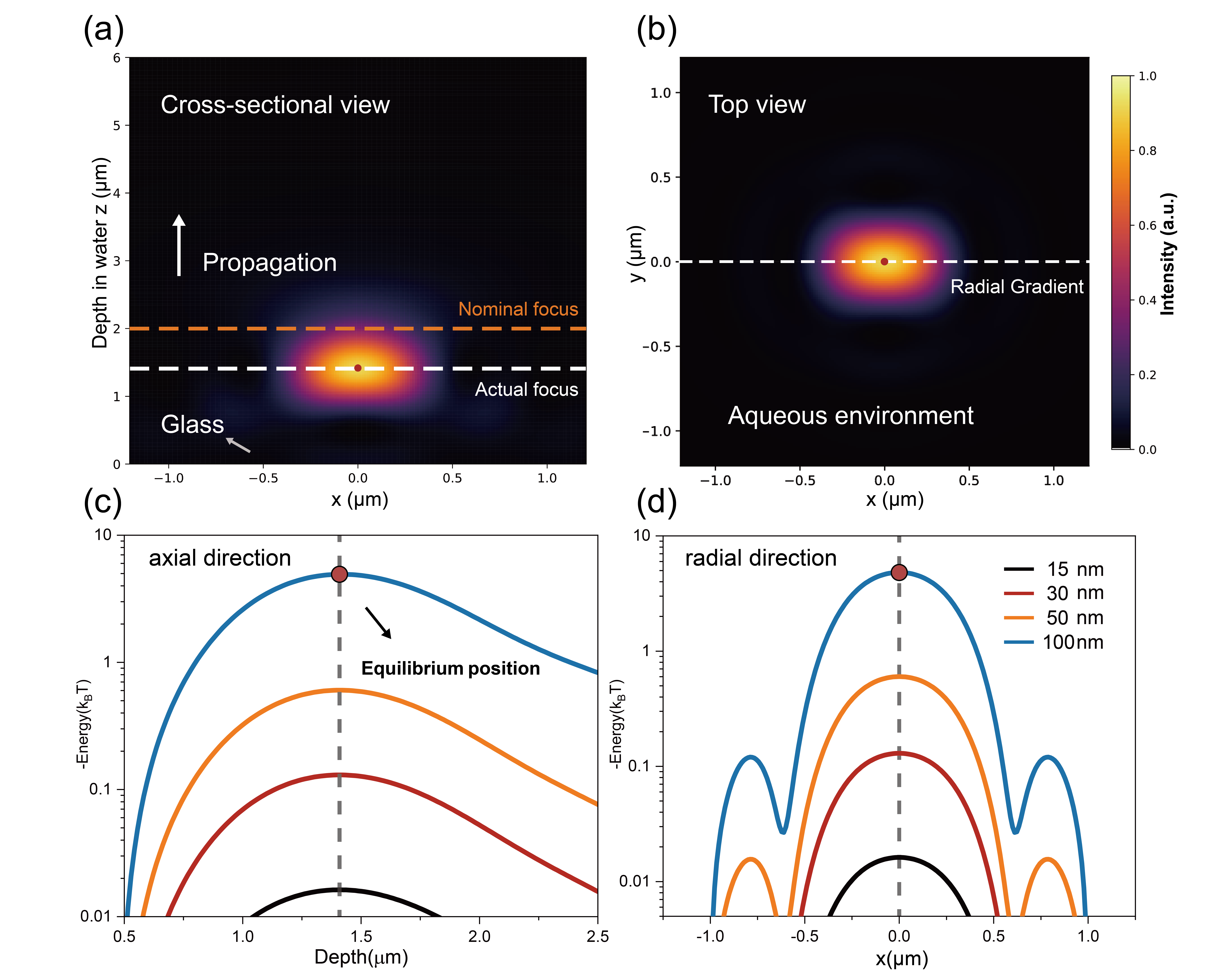}
\caption{Reconstructed trapping field and simulated optical trapping potentials at a nominal focal depth of $2~\mu\mathrm{m}$. (a) Cross-sectional view of the focal-field distribution across the glass--water interface, showing the nominal trapping plane and the aberration-shifted focal position. (b) Top view of the reconstructed focal field at the trapping plane. The slight asymmetry of the focal spot arises from the $x$-polarized trapping beam under high-NA focusing. (c) Axial and (d) radial potential-energy landscapes for particles of different diameters.}
\label{Pic_C}
\end{figure}

Figure~\ref{Pic_C} summarizes the reconstructed focal field and the corresponding trapping potentials under the selected trapping condition. Figure~\ref{Pic_C}(a) presents a cross-sectional view of the reconstructed field, with the beam propagating from the bottom glass substrate toward the aqueous region. This configuration corresponds to the sample region indicated by the orange dashed frame in Figure~\ref{Pic_B}. The lower red line indicates the glass--water interface. The orange dashed line marks the nominal trapping plane corresponding to the preset focal depth, while the white dashed line indicates the actual focal position after the spherical-aberration-induced shift. The focal position is shifted by approximately $0.584~\mu\mathrm{m}$. Figure~\ref{Pic_C}(b) presents the top view of the reconstructed optical field at the trapping plane. This asymmetry originates from the $x$-polarized trapping beam under high-NA focusing, which leads to different focal-field distributions along the $x$ and $y$ directions. The reconstructed focal field was used as the input for the subsequent optical-force calculation.

With the trapping field under the selected experimental condition established, we next evaluated whether it could provide sufficient optical confinement for nanoparticles of different sizes. The Rayleigh-force expressions introduced above are useful for illustrating the strong particle-size dependence of the optical force. However, these expressions are derived under the electric-dipole approximation and are strictly valid only when the particle dimensions are much smaller than the wavelength in the surrounding medium. As the particle diameter approaches approximately $100~\mathrm{nm}$, this condition becomes progressively less well satisfied. Moreover, while the gradient-force contribution scales as $r^3$, the Rayleigh scattering cross section scales as $r^6$, making the predicted scattering response more sensitive to particle size. Consequently, the Rayleigh approximation may no longer provide sufficient quantitative accuracy for evaluating the axial optical force in this size range. To obtain a more reliable optical-force calculation under the experimental conditions, the reconstructed focal field was therefore used as the input for the T-matrix calculation~\cite{nieminen2007optical,nieminen2011tmatrix}.

In the T-matrix calculation, a multipolar expansion was used to describe the interaction between the reconstructed incident field and the nanoparticle. The scattered-field coefficients were obtained from the incident-field coefficients through the particle T-matrix, and the time-averaged optical force was evaluated from the Maxwell stress tensor,
\begin{equation}
\langle\mathbf{F}\rangle =
\oint_S
\langle\mathbf{T}_{\mathrm{M}}\rangle
\cdot\hat{\mathbf{n}}\,
\mathrm{d}S.
\end{equation}
Here, $\mathbf{T}_{\mathrm{M}}$ represents the Maxwell stress tensor, and the integration is performed over a closed surface surrounding the nanoparticle. Radial and axial force profiles were calculated by translating the particle through the focal region. An additional consideration is the comparison between optical confinement and thermal fluctuations. Since the optical-force magnitude alone does not directly provide an energy scale for such a comparison, the calculated force profiles were converted into effective trapping potentials. The corresponding effective trapping potentials were obtained according to
\begin{equation}
U_i(q_i) =
-\int_{q_{i,0}}^{q_i}
F_i(q_i')\,\mathrm{d}q_i',
\qquad i=x,z.
\end{equation}
Here, $q_i$ denotes the particle displacement along the corresponding coordinate, $q_{i,0}$ represents the stable equilibrium position where the optical force vanishes, and $i=x,z$ correspond to the radial and axial directions, respectively. 

The calculated trapping potentials are shown in Figure~\ref{Pic_C}(c) and Figure~\ref{Pic_C}(d). The axial potential was calculated along the beam-propagation ($z$) direction, while the radial potential was evaluated along the $x$ direction of the reconstructed focal field. The potential profiles were obtained by integrating the calculated optical forces according to Eq.~(6). For visualization, the potentials are plotted as $-U/k_B T$, such that a larger positive peak corresponds to a deeper trapping potential. The stable trapping position corresponds to a local minimum of the actual potential energy and therefore appears as a maximum in the plotted curves. The energy difference between this minimum and the surrounding potential determines the resistance of the trapped particle to thermal fluctuations. A potential depth substantially larger than $k_B T$ therefore corresponds to more stable confinement.

The calculated potential depth increases strongly with particle diameter in both the axial and radial directions. For particles with dimensions comparable to bare CQDs, the optical potential remains relatively shallow and is comparable to or smaller than $k_B T$, making the confinement susceptible to thermal fluctuations. Increasing the effective particle diameter toward 100~nm substantially deepens the trapping potential at the same incident optical power. Compared with bare CQDs, the potential depth is increased by approximately three orders of magnitude, indicating a substantially more favorable confinement regime for the enlarged polymer-encapsulated particles.

Overall, the simulations first establish a preferred trapping-depth range by evaluating the influence of spherical aberration on the reconstructed focal field. Using the selected trapping condition, the calculated potential landscapes then show that enlarged nanoparticles exhibit substantially deeper radial and axial trapping wells, indicating stronger confinement against thermally driven fluctuations.

\subsection{2.4 Effective Optical Trapping of CQDs}

\begin{figure}[!ht]
\centering
\includegraphics[width=16cm]{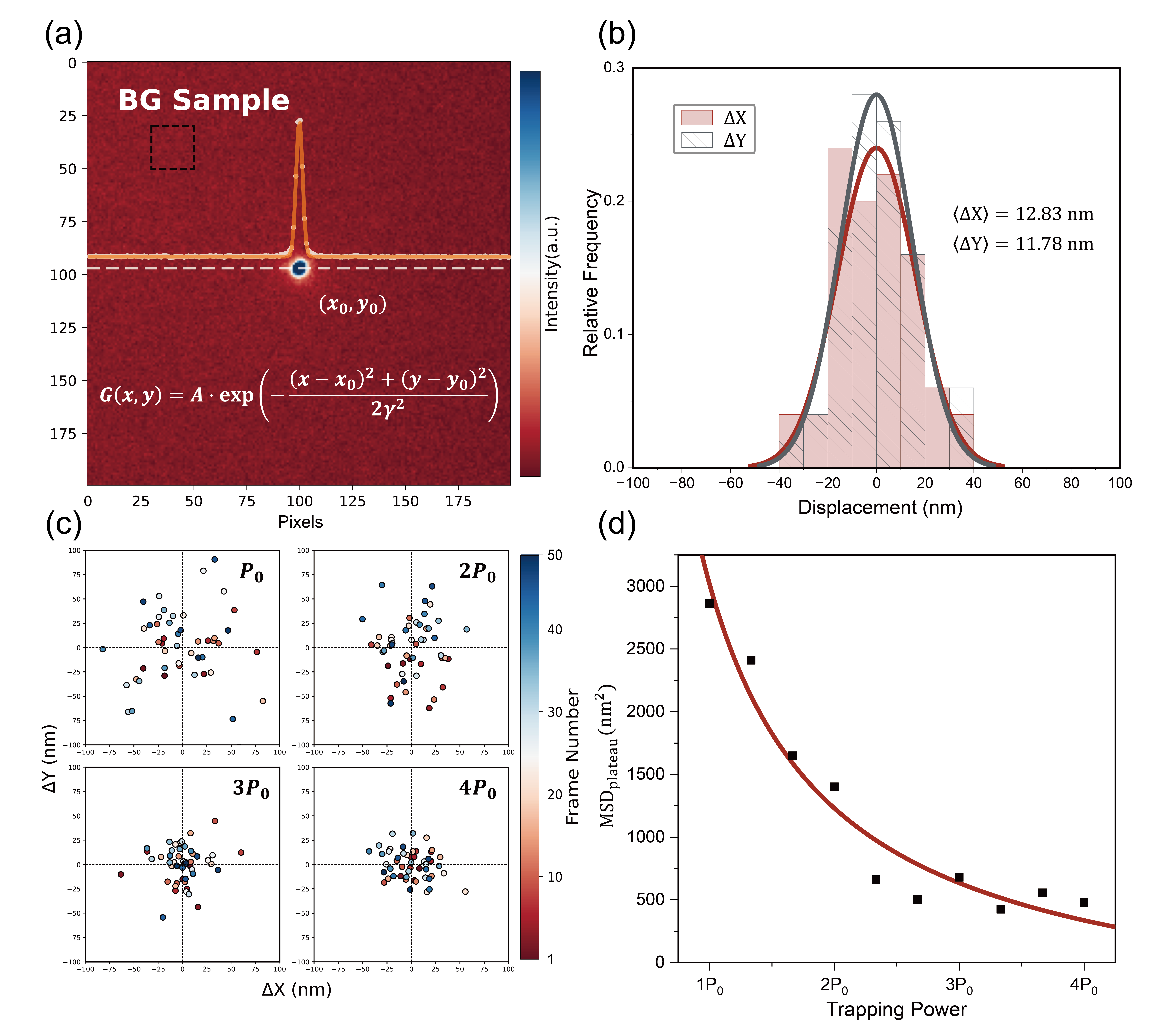}
\caption{Spatial confinement and trajectory analysis of an optically trapped PQD. 
(a) Representative wide-field PL image of a trapped PQD.
(b) Histograms of the positional deviations, with mean absolute deviations below 20~nm.
(c) Extracted $x$--$y$ positions at different trapping powers with an integration time of 100~ms per frame, showing progressively narrower distributions at higher powers.
(d) Plateau mean-square displacement (MSD) along one lateral direction, which decreases with increasing trapping power, indicating stronger optical confinement.}
\label{Pic_D}
\end{figure}

The trapping stability of the encapsulated nanoparticles was evaluated by trajectory tracking based on their photoluminescence signals recorded with an sCMOS camera. Owing to the diffraction limit, the physical boundary of an individual nanoparticle cannot be directly resolved with sufficient precision in the recorded images. The center position of each trapped PQD was therefore extracted by fitting with a two-dimensional Gaussian function \cite{novotny2006principles},
\begin{equation}
G(x,y)
=
A \cdot \exp\left[
-\frac{(x-x_0)^2+(y-y_0)^2}{2\gamma^2}
\right],
\end{equation}
where $A$ is the peak intensity, $(x_0,y_0)$ represents the fitted center position of the particle, and $\gamma$ describes the width of the Gaussian profile. A least-squares fitting procedure was used to determine the particle center. The fitted center coordinates were obtained frame by frame to reconstruct the particle trajectory. The resulting trajectory was then referenced to the mean particle position to quantify the positional fluctuations within the trap.

Figure~\ref{Pic_D} summarizes the spatial confinement and trajectory analysis of a representative trapped PQD nanoparticle. Figure~\ref{Pic_D}(a) presents a wide-field PL image of an individual PQD nanoparticle confined near the center of the trapping region, with a calibrated image pixel size of approximately 126~nm/pixel. The intensity profile above the particle image is extracted along the dashed line passing through the fitted particle center, together with the corresponding profile obtained from the two-dimensional Gaussian fit. The region enclosed by the black dashed box was selected as the background region, and its average intensity was used for background subtraction. The positional deviations extracted over the recorded frames are summarized by the histograms in Figure~\ref{Pic_D}(b), where $\Delta x=x-\langle x\rangle$ and $\Delta y=y-\langle y\rangle$ represent the deviations from the mean particle position along the $x$ and $y$ directions, respectively. These deviations provide a direct measure of the extent of the particle's positional fluctuations within the optical trap and are therefore used to evaluate the degree of spatial confinement\cite{kotlarchyk2010characterization}. The distributions of $\Delta x$ and $\Delta y$ are well approximated by Gaussian functions, as expected for a particle undergoing thermal fluctuations near the minimum of an approximately harmonic optical trapping potential. The mean absolute deviations, $\langle|\Delta x|\rangle$ and $\langle|\Delta y|\rangle$, remain below 20~nm, which is far below the sphere diameter used for our PQD design, indicating that the PQD remains well localized around its mean trapping position during the measurement.

The trajectory alone may not fully exclude effects such as transient surface sticking. To further verify that the observed localization originates from optical trapping, the position of the same PQD nanosphere was tracked at different trapping powers. Figure~\ref{Pic_D}(c) shows the extracted $x$--$y$ positions over consecutive frames. As the trapping power increases from $P_0$ to $4P_0$, where $P_0$ is approximately $29.40~\mathrm{mW}$ at the sample plane, the spatial distribution progressively narrows, indicating stronger confinement around the trapping center. 

The power-dependent positional fluctuations were further quantified using the mean-square displacement (MSD)\cite{pesce2020optical,grimm2012high}. Here, the MSD was evaluated along a single direction. For a Brownian particle confined in this potential, the MSD initially increases with the time interval and approaches a plateau once the observation time exceeds the characteristic relaxation time\cite{pesce2020optical}. This plateau reflects the finite range of positional fluctuations allowed by the trapping potential. In the present measurements, a lag time of 100~ms was sufficiently long for the MSD to reach this plateau. The plateau value, $\mathrm{MSD}_{\rm plateau}=\displaystyle\lim_{\tau\rightarrow\infty}\mathrm{MSD}(\tau)$, was therefore used as an effective measure of the positional fluctuations. As shown in Figure~\ref{Pic_D}(d), $\mathrm{MSD}_{\rm plateau}$ decreases monotonically with increasing trapping power, indicating reduced positional fluctuations and stronger confinement at higher trapping powers, consistent with the progressive narrowing of the positional distributions in Figure~\ref{Pic_D}(c).

Overall, these measurements demonstrate stable optical confinement of the encapsulated PQDs under the experimental conditions. The observed localization, together with its systematic dependence on trapping power, agrees with the calculated trapping behavior and confirms that the particles are effectively confined by the optical trap.

\subsection{2.5 Quantum-Optical Measurement of Spatially Confined CQDs}

\begin{figure}[!ht]
\centering
\includegraphics[width=16cm]{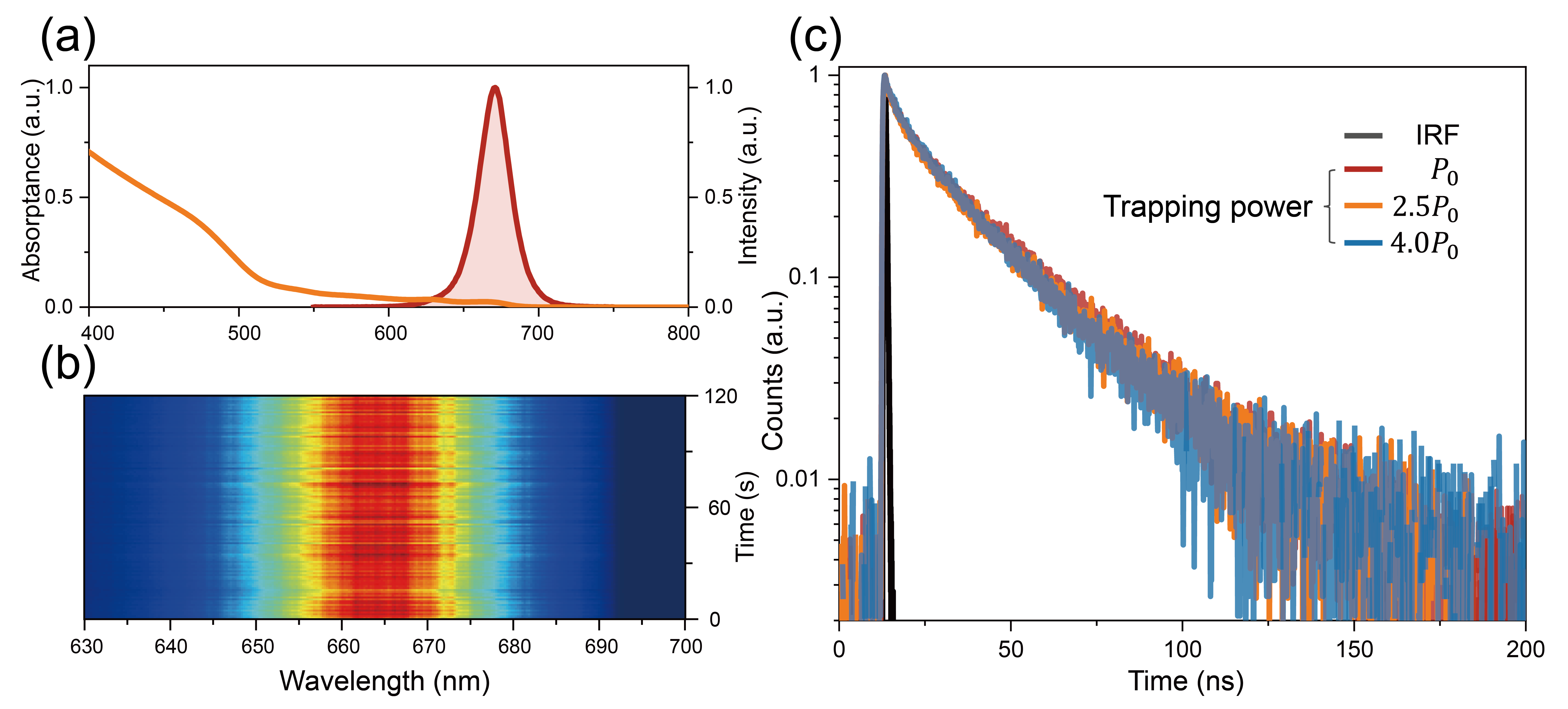}
\caption{
Optical characterization of PQDs under optical trapping.
(a) Absorption and photoluminescence spectra of the CQD ensemble, showing a pronounced decrease in absorption at longer wavelengths.
(b) Emission spectra of a single trapped PQD nanoparticle recorded over 120~s.
(c) Photoluminescence decay curves of a trapped PQD measured at different trapping powers.}
\label{Pic_E}
\end{figure}

Having established stable optical confinement, we subsequently investigate the impact of trapping conditions on the emission and quantum properties of the encapsulated PQDs. To contextualize the direct evaluation of trapped emission, we first analyzed the ensemble optical spectra to determine the spectral separation between the CQD absorption band and the 1064 nm trapping wavelength. As shown in Figure \ref{Pic_E}(a), the absorption spectrum exhibits a rapid decline at longer wavelengths, approaching the baseline near the 800 nm measurement limit. Consequently, the photon energy at 1064 nm lies well below the band-edge transition of the CQDs, suggesting that direct photon absorption is negligible and validating the suitability of this wavelength for optical trapping with minimal direct excitation.

The spectral stability of the trapped PQD was then examined during continuous optical confinement. Figure~\ref{Pic_E}(b) shows the evolution of the emission spectrum of an individual trapped PQD over a continuous 120~s measurement. The spectral profiles remain highly consistent throughout the measurement, with no appreciable shift of the emission peak or systematic decrease in emission intensity. Because changes in the emission dynamics may occur without producing an obvious modification of the steady-state spectrum, the PL decay was further measured to evaluate the influence of the trapping field on the emission process. Figure~\ref{Pic_E}(c) shows the PL decay of the trapped PQD at different trapping powers. Across a fourfold range of trapping power, the decay profiles remain closely overlapping, with no systematic acceleration of the decay or emergence of an additional fast component. 

\begin{figure}[!ht]
\centering\includegraphics[width=10cm]{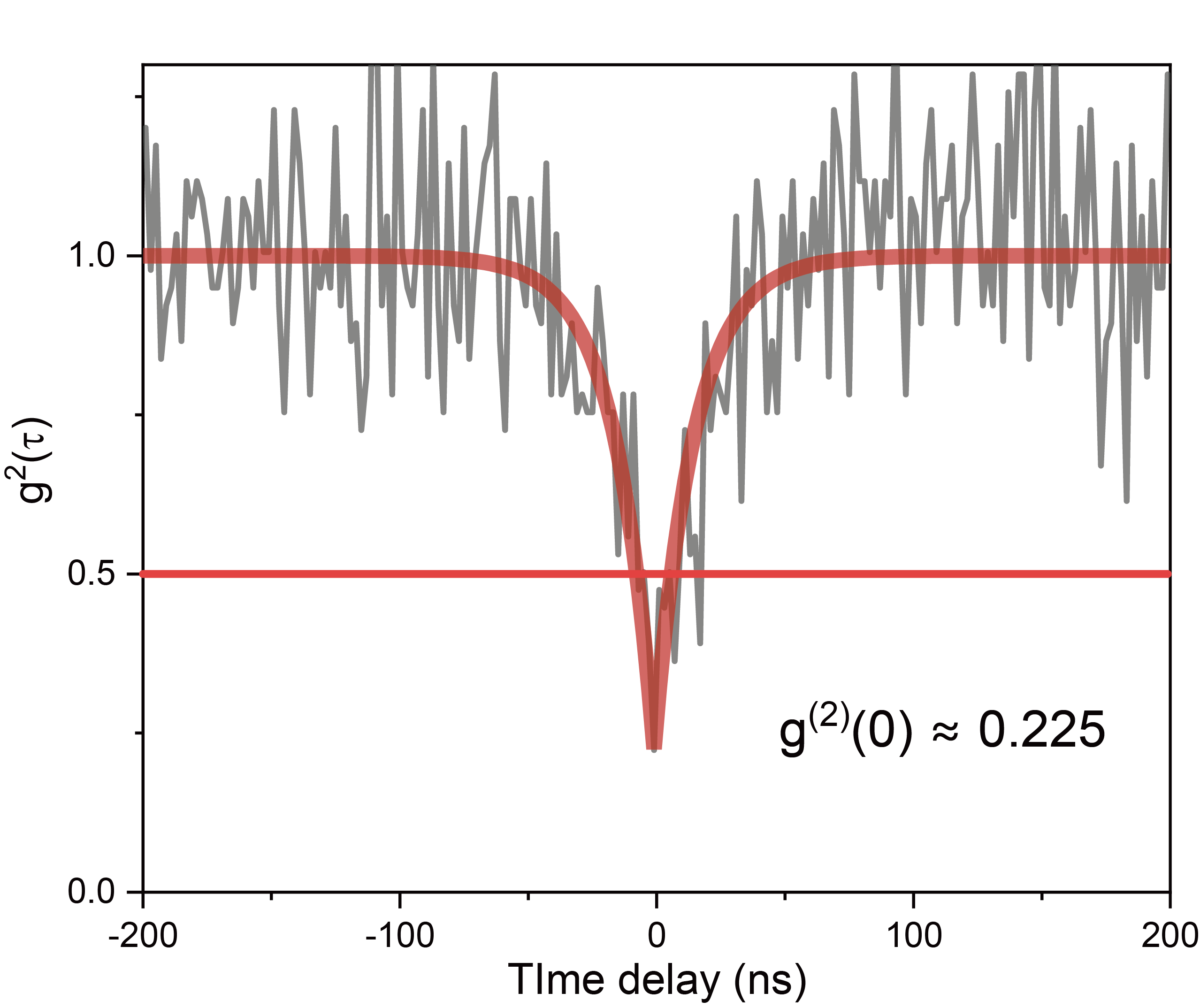}
\caption{
Single-photon characterization of a substrate-immobilized PQD under trapping-laser illumination. The measured $g^{(2)}(0)$ is approximately 0.225, demonstrating antibunched emission.}
\label{Pic_F}
\end{figure}

For a quantum emitter, preservation of the emission spectrum and decay dynamics does not by itself establish that the single-photon emission characteristic remains compatible with the trapping-laser environment. Single-photon emission is commonly evaluated through the second-order photon-correlation function $g^{(2)}(\tau)$, which describes the temporal correlation between successive photon-detection events. In particular, $g^{(2)}(0)$ characterizes the relative probability of detecting two photons at zero time delay\cite{fox2006quantum}. For an ideal single-photon emitter, simultaneous photon emission is suppressed, producing photon antibunching around zero time delay, while $g^{(2)}(0)<0.5$ is commonly used as an experimental criterion for single-photon emission\cite{lin2017electrically}. We therefore performed a photon-correlation measurement on an individual PQD containing a single quantum dot, with the PQD immobilized on a substrate and exposed to the focused 1064~nm trapping laser at a power of approximately $3P_0$. At this trapping power, the reduced MSD shown in Figure~\ref{Pic_D}(d) indicates that an individual encapsulated CQD sphere can be well confined under the same optical trapping condition. As shown in Figure~\ref{Pic_F}, a pronounced antibunching dip is observed around zero time delay, with a fitted value of $g^{(2)}(0)\approx0.225$. The observed antibunching demonstrates that single-photon emission remains observable under 1064~nm trapping-laser illumination. Although the photon-correlation measurement was not performed on a freely trapped PQD, it provides a direct verification that the optical environment associated with the trapping laser remains compatible with single-photon emission.

Taken together, these measurements show that optical confinement can be achieved without introducing appreciable changes to the spectral response or photoluminescence decay of the encapsulated PQDs. The preservation of pronounced antibunching under 1064~nm trapping-laser illumination further demonstrates that their single-photon emission remains compatible with the optical trapping environment. These results establish the feasibility of combining spatial confinement with quantum-optical measurements of colloidal quantum emitters in an optical trapping environment.

\section{3. Conclusion}
In conclusion, we demonstrate stable optical confinement of polymer-encapsulated colloidal quantum dots while preserving their quantum-optical characteristics under the investigated trapping conditions. By increasing the effective particle size through polymer encapsulation, the optical trapping force is substantially enhanced, enabling reliable spatial localization against Brownian motion. The trapped PQDs exhibit stable emission spectra and photoluminescence decay behavior over the investigated trapping-power range, while photon-correlation measurements further verify that pronounced antibunching remains compatible with the 1064~nm trapping-laser environment. These results establish a practical route toward bringing spatial control and quantum-optical characterization together for colloidal quantum emitters, providing a basis for their controlled spatial organization into functional quantum assemblies.

\begin{acknowledgement}

Z.N. thanks Prof. Xiao-Gang Peng and all the members of his research group in the Department of Chemistry, Zhejiang University for helpful discussions in chemical synthesis.
This work is supported by the Zhejiang Province Leading Geese Plan (2024C01105), the National Future Industry Innovation Mission, the Beijing Natural Science Foundation (L248103), the National Key Research and Development Program of China (2021YEB2800500), and the National Natural Science Foundation of China (61574138, 61974131).
\end{acknowledgement}

\section{Author Contributions}
Z.N., C.J., and X.L. conceived the idea. J.L. synthesized the high-quality CQDs. Z.N., X.L., W.F., X.C., J.Y., Y.W., D.H., Y.L., and C.J. contributed to the preparation of the experimental setup. Z.N. conducted all experiments. C.J. and X.L. supervised the project. C.J. directed the project. Z.N. and C.J. wrote the manuscript with contributions from all authors.

\section{Competing Interests}
The authors declare no competing interests.
\newpage
\section{Table of Contents (TOC) Graphic}
\begin{figure}[htbp]
\centering\includegraphics[width=16cm]{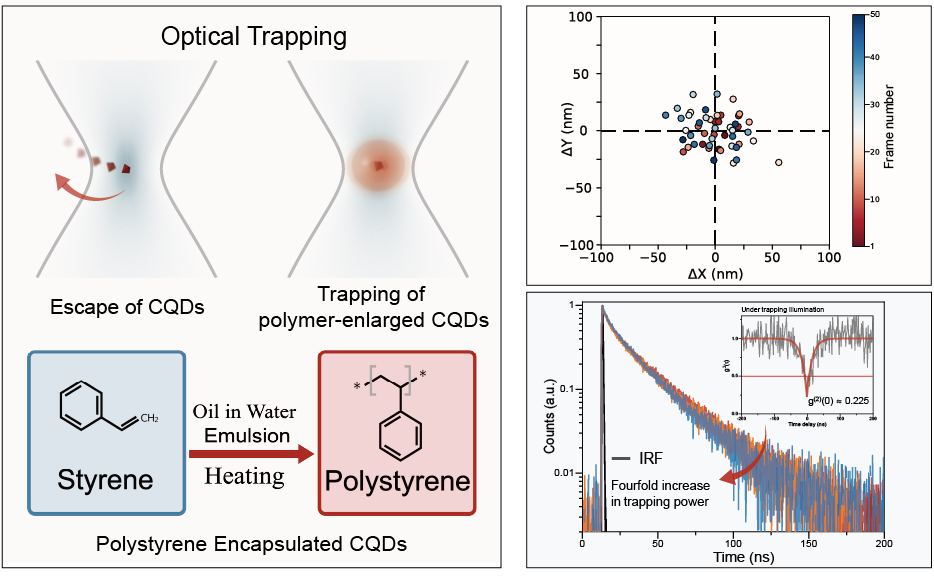}
\caption{
TOC.}
\label{Pic_Toc}
\end{figure}

\newpage
\bibliography{achemso-demo}

\end{document}


\newpage
\subsection{1. Three-dimensional Focal-field Reconstruction}
The three-dimensional focal field of the 1064~nm trapping beam was reconstructed using a vectorial Debye model for high-NA focusing through the glass--water interface \cite{torok1995electromagnetic, torok1997electromagnetic}. The calculation accounts for Fresnel transmission, polarization transformation, and the spherical-aberration phase introduced by the refractive-index mismatch. The refractive indices of the glass/oil side and water were taken as $n_1=1.518$ and $n_2=1.33$, respectively, and the numerical aperture of the objective was set to 1.42.

In the vectorial Debye formalism, the incident beam is represented as a collection of plane-wave components propagating at different angles. After transmission through the glass--water interface, these components are coherently summed to reconstruct the focused field. For each incident angular component characterized by the angle $\theta_1$, the corresponding propagation angle $\theta_2$ in water is determined from Snell's law,

\begin{equation}
n_1\sin\theta_1=n_2\sin\theta_2.
\end{equation}

The maximum convergence angle defined by the objective is

\begin{equation}
\alpha=\sin^{-1}\left(\frac{\mathrm{NA}}{n_1}\right).
\end{equation}

Because $n_1>n_2$, only angular components that remain propagating in the aqueous medium were considered. The upper limit of the angular integration was therefore taken as

\begin{equation}
\theta_{\max}
=
\min\left(\alpha,\theta_c\right),
\qquad
\theta_c=\sin^{-1}\left(\frac{n_2}{n_1}\right),
\end{equation}
where $\theta_c$ is the critical angle of the glass--water interface.

The $s$- and $p$-polarized components of each angular component are transmitted differently across the interface. Following the formulation of T\"or\"ok \textit{et al.}, the corresponding Fresnel transmission coefficients are written as

\begin{equation}
t_s=
\frac{2\sin\theta_2\cos\theta_1}
{\sin(\theta_1+\theta_2)},
\end{equation}
and

\begin{equation}
t_p=
\frac{2\sin\theta_2\cos\theta_1}
{\sin(\theta_1+\theta_2)\cos(\theta_1-\theta_2)}.
\end{equation}

Under the Abbe sine condition, each angular component is weighted by an angle-dependent factor that accounts for both the high-NA apodization and the spherical-coordinate integration measure. The combined angular weighting factor is therefore written as

\begin{equation}
A(\theta_1)=\sqrt{\cos\theta_1}\sin\theta_1,
\end{equation}
where the factor $\sqrt{\cos\theta_1}$ originates from the apodization associated with high-NA focusing, while $\sin\theta_1$ accounts for the angular weighting associated with integration over different propagation directions.

For an $x$-polarized incident field, the transmitted angular spectrum can then be decomposed into three Debye-type integrals,

\begin{equation}
I_0(\rho,h)
=
\int_0^{\theta_{\max}}
A(\theta_1)
\left(t_s+t_p\cos\theta_2\right)
J_0(k_0n_1\rho\sin\theta_1)
e^{i\Phi(\theta_1,h)}
\,d\theta_1 ,
\end{equation}

\begin{equation}
I_1(\rho,h)
=
\int_0^{\theta_{\max}}
A(\theta_1)
t_p\sin\theta_2
J_1(k_0n_1\rho\sin\theta_1)
e^{i\Phi(\theta_1,h)}
\,d\theta_1 ,
\end{equation}
and

\begin{equation}
I_2(\rho,h)
=
\int_0^{\theta_{\max}}
A(\theta_1)
\left(t_s-t_p\cos\theta_2\right)
J_2(k_0n_1\rho\sin\theta_1)
e^{i\Phi(\theta_1,h)}
\,d\theta_1 ,
\end{equation}
where $J_0$, $J_1$, and $J_2$ are Bessel functions of the first kind, $k_0=2\pi/\lambda_0$, and $\rho$ denotes the radial coordinate in the plane.

The refractive-index mismatch also introduces an angle-dependent propagation phase. In the present geometry, the phase term is written as

\begin{equation}
\Phi(\theta_1,h)
=
k_0
\left[
n_2h\cos\theta_2
-
n_1d\cos\theta_1
\right],
\end{equation}
where $d$ is the nominal focal depth measured from the glass--water interface and $h$ denotes the physical depth in the aqueous region. Because different angular components accumulate different phases after transmission through the interface, the refractive-index mismatch introduces spherical aberration, thereby shifting the actual focal position and modifying the focal-field distribution.

The three Cartesian components of the focal electric field are reconstructed from the Debye integrals as

\begin{equation}
E_x=-i\left[I_0+I_2\cos(2\phi)\right],
\end{equation}

\begin{equation}
E_y=-iI_2\sin(2\phi),
\end{equation}
and

\begin{equation}
E_z=-2I_1\cos\phi.
\end{equation}
Here, $\phi$ denotes the azimuthal angle in the radial plane, defined together with the radial coordinate $\rho$ by

\begin{equation}
x=\rho\cos\phi,
\qquad
y=\rho\sin\phi,
\qquad
\rho=\sqrt{x^2+y^2}.
\end{equation}

The three-dimensional focal-field intensity is consequently obtained from

\begin{equation}
I(x,y,h)
\propto
|E_x|^2+|E_y|^2+|E_z|^2.
\end{equation}

\begin{figure}[htbp]
\centering
\includegraphics[width=14cm]{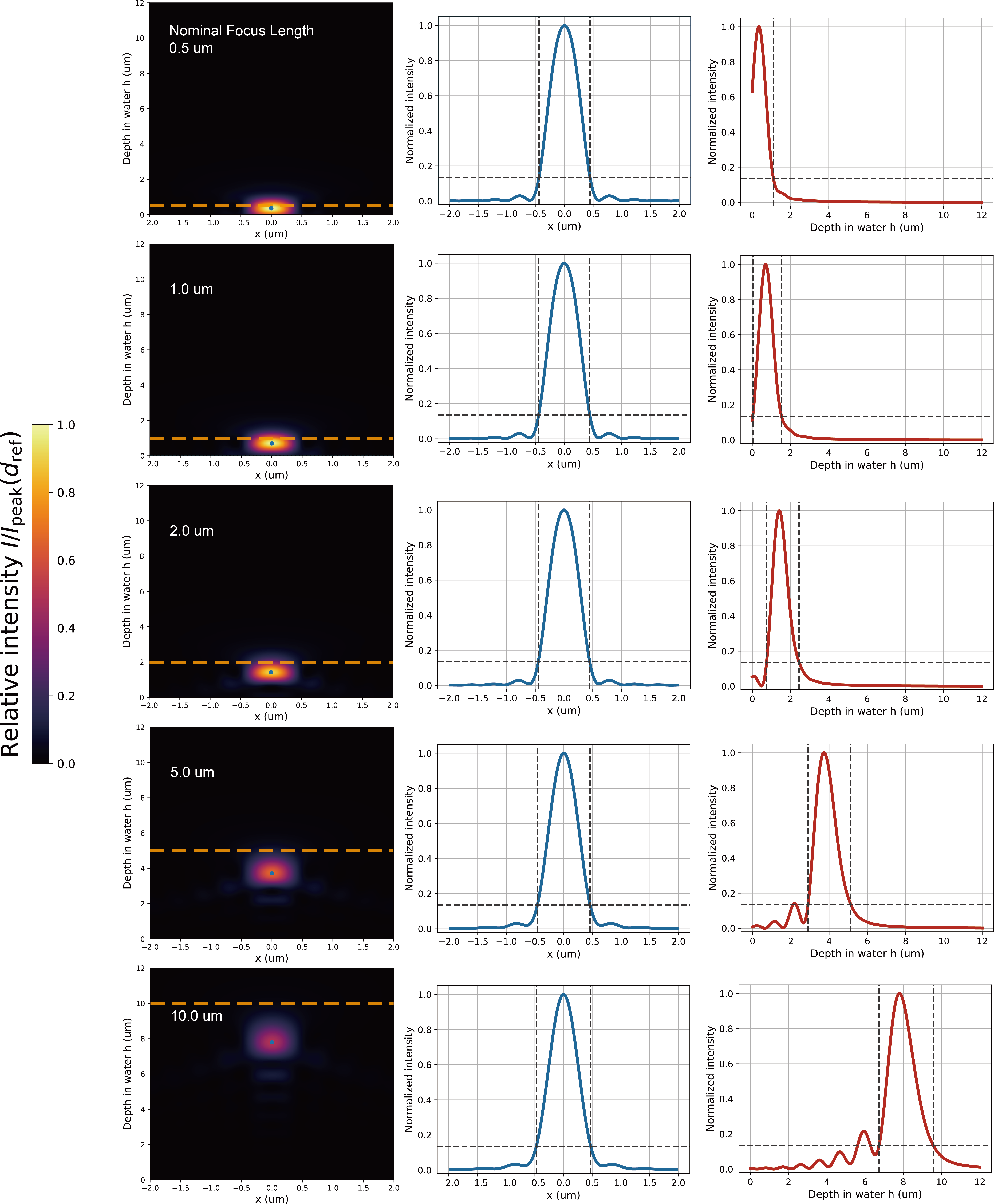}
\caption{Evolution of the reconstructed 1064~nm trapping field with nominal focal depth. The left column shows the focal-field distributions in the $x$--$h$ plane at different nominal focal depths, where the dashed horizontal lines indicate the corresponding nominal focal planes. The middle column shows the corresponding radial intensity profiles through the focal maximum, where the radial focal width changes comparatively little with focal depth. The right column shows the axial intensity profiles along the beam-propagation direction, exhibiting pronounced axial broadening and focal-position shift.}
\label{S1}
\end{figure}

By evaluating the above field components over the spatial coordinates $(x,y,h)$, the three-dimensional vectorial focal-field distribution of the trapping beam is reconstructed. The $x$--$h$ cross-sectional distribution and the $x$--$y$ distribution at the trapping plane are subsequently extracted for visualization and analysis.

The focal field was evaluated for nominal focal depths ranging from $0.5$ to $10~\mu\mathrm{m}$. For each nominal depth, the actual focal position was identified from the maximum of the reconstructed intensity distribution. The radial and axial focal-field widths were then determined using the $1/e^2$ intensity criterion. The radial width is expressed as the waist diameter $2w_0$, whereas the axial width is defined along the beam-propagation direction. The peak intensity at each focal depth was normalized to that obtained at $d=0.5~\mu\mathrm{m}$.

Figure~\ref{S1} shows the reconstructed focal-field distributions and the corresponding radial and axial intensity profiles at representative nominal focal depths. With increasing focal depth, the focal field becomes progressively broadened along the beam-propagation direction, accompanied by an increasing displacement of the intensity maximum from the nominal focal plane. In comparison, the radial intensity profile changes relatively little over the investigated focal-depth range. The axial profiles further show increasingly pronounced distortion of the focal field at larger focal depths.

\begin{figure}[htbp]
\centering
\includegraphics[width=10cm]{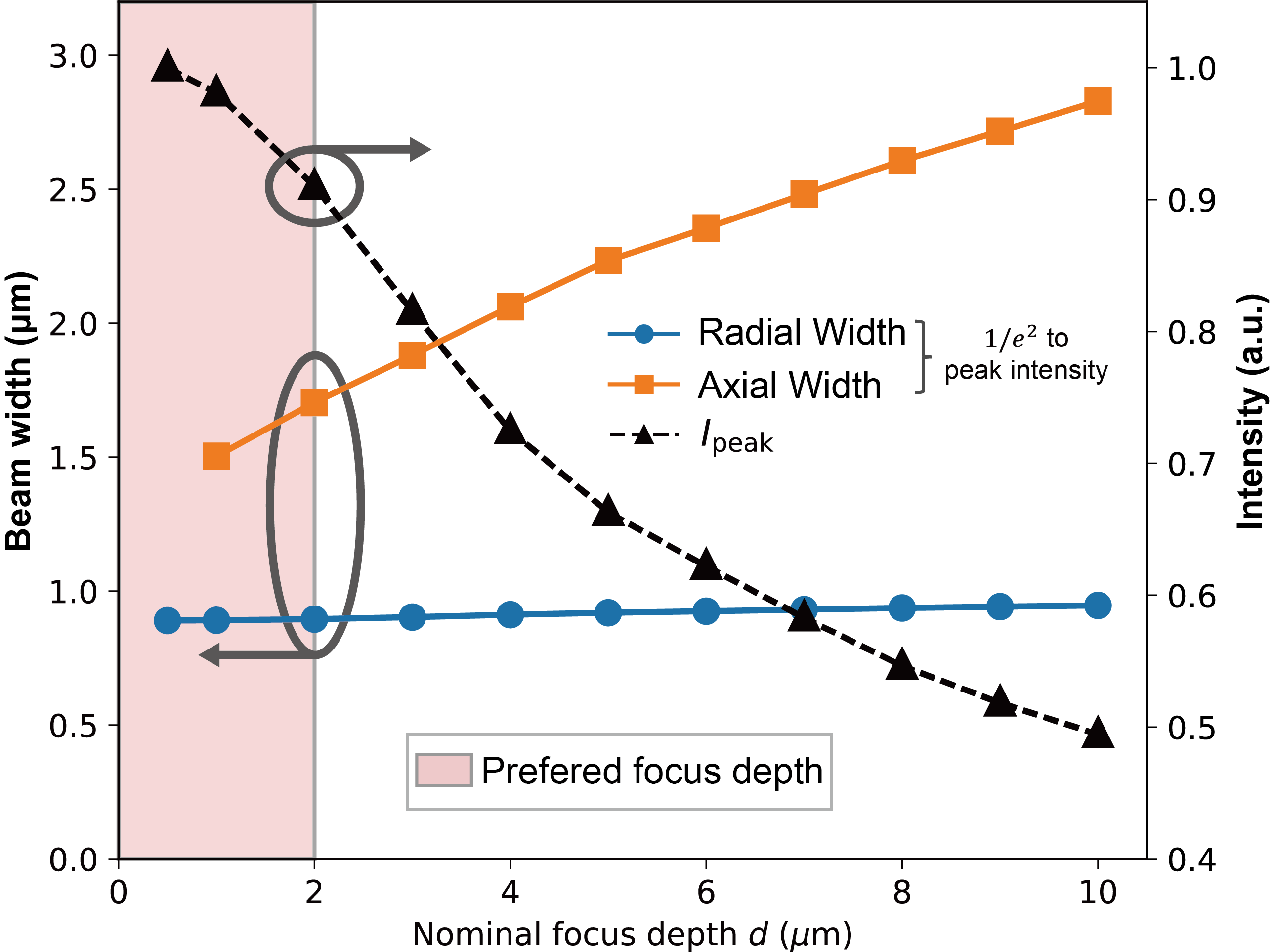}
\caption{Dependence of the reconstructed 1064~nm trapping field on the nominal focal depth $d$. The radial waist diameter $2w_0$, axial $1/e^2$ width, and normalized peak intensity $I_{\mathrm{peak}}/I_{\mathrm{ref}}$ are shown as functions of focal depth. The increasing axial width and decreasing peak intensity reflect the progressively stronger influence of spherical aberration at larger focal depths. A nominal trapping depth below approximately $2~\mu\mathrm{m}$ is therefore preferable for maintaining a sufficiently confined trapping field, as indicated by the left red-shaded region.}
\label{S2}
\end{figure}

These variations are quantitatively summarized in Figure~\ref{S2}. As the nominal focal depth increases from $0.5~\mu\mathrm{m}$ to $10~\mu\mathrm{m}$, the radial waist diameter changes only slightly, whereas the axial width increases by approximately a factor of two. Meanwhile, the peak intensity decreases by approximately 50\% relative to that at a focal depth of $0.5~\mu\mathrm{m}$. These results indicate that increasing the focal depth progressively weakens the axial confinement of the focal field because of the increasing spherical aberration at the glass--water interface. Since the optical gradient force depends on the spatial variation of the intensity, a nominal trapping depth below approximately $2~\mu\mathrm{m}$ is therefore preferable for maintaining a sufficiently confined trapping field, as indicated by the left red-shaded region.

\subsection{2. Elements of Optical Setup}
Optical characterization was conducted using a custom-built micro-photoluminescence ($\mu$-PL) system. The PQD nanoparticles were excited using either a continuous-wave (CW) or a picosecond pulsed laser operating near 450~nm, while optical trapping was provided by a 1064~nm polarization-maintaining single-mode fiber laser (Connect CoSF-D-YB-B) with a maximum output power of 10~W. The trapping beam was expanded by a beam expansion system before entering the objective to provide sufficient filling of the back aperture. A dichroic mirror was used to separate the fluorescence signal from the trapping and excitation beams. A high-numerical-aperture oil-immersion objective (Olympus PlanApo N, 60$\times$, NA 1.42) was used for optical trapping, excitation, and fluorescence collection. 
Imaging and trajectory measurements were performed using an sCMOS camera (Tucsen Dhyana 95V2). The minimum achievable effective spatial sampling of the imaging system is 57.13~nm per pixel. The emitted photons were detected using single-photon avalanche photodiodes (SPADs, PerkinElmer SPCM-AQR-14), while fluorescence lifetime and photon-correlation measurements were recorded using a time-correlated single-photon counting (TCSPC) module (PicoQuant HydraHarp 400). 

\subsection{3. Materials and Sample Cell Preparation}
The CdSe/CdS/ZnS CQDs used in this work were purchased from NajingTech and subsequently encapsulated in polystyrene to form PQD nanoparticles. A similar PQD preparation procedure has been described previously~\cite{ni2026quantum}.

The sample cell was fabricated from two commercially available glass coverslips with a thickness of approximately 170~$\mu$m. A shallow recessed region was formed in one of the coverslips to serve as the liquid sample chamber. The pattern of the recessed region was first defined by direct-write lithography (DWL) using APR~5350 photoresist, followed by etching with buffered oxide etchant (BOE) to a depth of approximately 2~$\mu$m. After removal of the resist and cleaning, the etched coverslip was assembled with a second coverslip to form a sealed shallow liquid cell. The PQD suspension was confined within the approximately 2~$\mu$m-deep chamber between the two glass surfaces. During optical trapping measurements, the etched coverslip was placed on the objective side, allowing the trapping beam to be focused into the shallow liquid layer through the high-NA oil-immersion objective.

\newpage

\bibliography{achemso-demo}